\documentclass[conference,letterpaper]{IEEEtran}
\usepackage[letterpaper, left=0.625in, right=0.625in, bottom=1in, top=0.75in]{geometry}
\IEEEoverridecommandlockouts
\usepackage{cite}
\usepackage{amsmath,amssymb,amsfonts}
\usepackage{graphicx}
\usepackage{textcomp}
\usepackage{xcolor}
\def\BibTeX{{\rm B\kern-.05em{\sc i\kern-.025em b}\kern-.08em
    T\kern-.1667em\lower.7ex\hbox{E}\kern-.125emX}}

\usepackage[frozencache ,cachedir=.]{minted}

\usepackage{svg}
\usepackage{minted}
\usepackage{tcolorbox}
\usepackage{enumitem}
\usepackage{lipsum}
\usepackage{xcolor, soul}
\usepackage{float}
\usepackage{dblfloatfix}

\usepackage{algorithm} 
\usepackage[noend]{algpseudocode}

\usepackage{multicol}
\usepackage{multirow}
\usepackage{graphicx}
\usepackage{float}

\makeatletter
\newcommand{\newlineauthors}{%
  \end{@IEEEauthorhalign}\hfill\mbox{}\par
  \mbox{}\hfill\begin{@IEEEauthorhalign}
}
\makeatother

\begin{document}

\title{
ByteStack-ID: Integrated Stacked Model Leveraging Payload Byte Frequency for Grayscale Image-based Network Intrusion Detection
}


\makeatletter
\patchcmd{\@maketitle}
  {\addvspace{0.5\baselineskip}\egroup}
  {\addvspace{-1.0\baselineskip}\egroup}
  {}
  {}
\makeatother

\author{
\IEEEauthorblockN{Irfan Khan}
\IEEEauthorblockA{
\textit{Department of Marine Engineering Technology, in joint appointment with}\\ \textit{Electrical and Computer Engineering Department}\\ \textit{and  Computer Science \& Engineering Department,}\\ \textit{Texas A\&M University } \\
Galveston, TX, USA \\
irfankhan@tamu.edu}
\newlineauthors

\IEEEauthorblockN{Yasir Ali Farrukh}
\IEEEauthorblockA{\textit{Clean and Resilient Energy Systems (CARES) Lab}\\
\textit{Texas A\&M University} \\
College Station, TX, USA \\
yasir.ali@tamu.edu}
\and
\IEEEauthorblockN{Syed Wali}
\IEEEauthorblockA{\textit{Clean and Resilient Energy Systems (CARES) Lab}\\
\textit{Texas A\&M University} \\
College Station, TX, USA \\
syedwali@tamu.edu}\\
}

\maketitle

\begin{abstract}
In the ever-evolving realm of network security, the swift and accurate identification of diverse attack classes within network traffic is of paramount importance. This paper introduces "ByteStack-ID," a pioneering approach tailored for packet-level intrusion detection. At its core, ByteStack-ID leverages grayscale images generated from the frequency distributions of payload data, a groundbreaking technique that greatly enhances the model's ability to discern intricate data patterns. Notably, our approach is exclusively grounded in packet-level information, a departure from conventional Network Intrusion Detection Systems (NIDS) that predominantly rely on flow-based data. While building upon the fundamental concept of stacking methodology, ByteStack-ID diverges from traditional stacking approaches. It seamlessly integrates additional meta learner layers into the concatenated base learners, creating a highly optimized, unified model. Empirical results unequivocally confirm the outstanding effectiveness of the ByteStack-ID framework, consistently outperforming baseline models and state-of-the-art approaches across pivotal performance metrics, including precision, recall, and F1-score. Impressively, our proposed approach achieves an exceptional 81\% macro F1-score in multiclass classification tasks. In a landscape marked by the continuous evolution of network threats, ByteStack-ID emerges as a robust and versatile security solution, relying solely on packet-level information extracted from network traffic data.
\end{abstract}

\begin{IEEEkeywords}
Network Intrusion Detection, Stacking, Machine Learning, Cybersecurity, Internet of things.
\end{IEEEkeywords}


\section{Introduction}

The 21st century has ushered in an era of unprecedented era of technological advancement, reshaping our lives through rapid progress in computation and communication technologies \cite{10356319}. However, this rapid advancement has also expanded the cyber-threat landscape, emphasizing the paramount importance of security in our increasingly interconnected digital world \cite{cite4,covert}. Furthermore, technologies like the Internet of Things (IoT), where a myriad of low-cost, resource-constrained devices converge on the global stage, further exemplify this trend with an estimated 24.1 billion connected devices projected by 2030, compared to 500 million in 2003 \cite{hatton2020iot}. IoT's transformative power is evident as it revolutionizes how people and devices interact within connected ecosystems \cite{wali2021explainable}.

In this interconnected society, communication and information systems underpin critical infrastructure \cite{ghadi2023improved} and personal devices. IoT's extensive deployment, encompassing nearly 30\% of attacks in communication networks \cite{cite5}, poses significant security challenges due to diverse architectures and communication channels \cite{cite4}. Network Intrusion Detection Systems (NIDS) play a crucial role in identifying and mitigating cyberattacks within a layered defense strategy. Recent advancements emphasize automated cyberattack identification, with machine learning (ML) algorithms gaining traction \cite{farrukh2021sequential,khan4572172radiant}. In ML, ensemble methods are prominent \cite{gupta2020improving,mehdi2023machine}, harnessing the collective power of multiple models like Bagging, Boosting, and Stacking to enhance accuracy \cite{syarif2012application}.

This paper introduces "ByteStack-ID," an innovative approach employing an integrated stacking model that leverages network traffic payload information as grayscale images for IoT intrusion detection. Unlike flow-based approaches, ByteStack-ID operates at the packet level, inspecting packet payloads to reveal malicious network flows. While flow-based methods focus on lower-level threats within the TCP/IP protocol stack, potentially overlooking higher-level threats \cite{Lee2001RealDetection}, packet-based approaches, such as ByteStack-ID, offer a more comprehensive solution. Our contributions in this paper encompass:

\begin{enumerate}
\item Introduction of an innovative integrated stacking approach, featuring deep concatenated 2D CNN models.
\item Development of a novel method for transforming high-dimensional packet payload data into grayscale images based on byte frequency.
\item Conducting a comprehensive quantitative analysis that compares our proposed method not only with state-of-the-art approaches but also with base models.
\end{enumerate}

\section{Related Work}
While extensive research has been conducted in the domain of NIDS, a predominant focus has been on flow-based network traffic analysis \cite{farrukh4635437ais}. While some approaches have delved into packet-level data, addressing the challenge of handling the extensive feature space inherent in packet payloads, we have narrowed our review to such studies that specifically leverage packet-level information.

In packet-level intrusion detection, Wang et al. introduced the HAST-II architecture, employing deep Convolutional Neural Networks (CNNs) and Long Short-Term Memory (LSTM) networks to convert packets into two-dimensional grayscale images, enhancing traffic classification \cite{wang2017hast}. Likewise, Zhang et al. presented a multi-layer IDS model, incorporating CNNs and gcForest. Their novel P-Zigzag algorithm converts raw packet-based network traffic into grayscale images, facilitating initial detection and further classification \cite{zhang2019multiple}. Furthermore, the authors in \cite{AEIDS}, proposed an LSTM-based autoencoder through exploiting payloads and calculating reconstruction errors.

Regarding deep learning techniques using packets as text input, Liu et al. introduced PL-CNN and PL-RNN models. PL-RNN employs the LSTM network, trained on the first $n$ characters of payloads \cite{liu2019cnn}. Packet2Vec, on the other hand, employs Word2Vec, a shallow neural network, to generate packet vectors. N-grams are utilized to create a dictionary, and packet vectors, with a length of 128, serve as features for classification using Random Forest (RF) \cite{goodman2020packet2vec}. Extending the Word2Vec concept, Hassan et al. proposed PayloadEmbeddings, utilizing vector representations of bytes in network packet payloads for anomaly detection with a k-Nearest Neighbor (kNN) classifier \cite{Hassan2022IntrusionEmbeddings}. Vidal et al. introduced EsPADA, employing the n-gram technique ($n = 3$) to extract features from payloads and classify packets based on global similarity scores (GSCs) \cite{vidal2020espada}. However, it is imperative to acknowledge that the performance of NIDS at the packet level remains an area in need of further enhancement, particularly in the methods used to transform packet-level information into effective features. This underscores the pressing need for continued progress and innovation in this field.

\section{Methodology}

This section provides an overview of our methodology, covering the dataset used, preprocessing steps including converting data into grayscale images, and the training of our integrated stacked model, detailing its architecture and workflow.

\subsection{Dataset and Preprocessing}

There are several publicly available dataset within the NIDS field, yet a significant proportion of these datasets predominantly center around header information from network packets, emphasizing flow-based network traffic analysis. Unfortunately, datasets encompassing actual payload data are sparse, primarily due to privacy-related concerns. Additionally, some datasets that do incorporate packet data are often found in a raw, unlabeled format, making them less accessible for ongoing research.

For our study, we utilized the CIC-IDS2017 dataset \cite{Sharafaldin2018TowardCharacterization}, which comprises simulated packet-based and bidirectional flow-based network traffic, encompassing modern attacks and benign traffic. The dataset is available in two formats: the original packet capture (PCAP) files (packet-based data) and CSV files (flow-based data). Since we focused primarily on packet-based network traffic, we opted to work with the raw packets from CIC-IDS2017.

Working with raw packet data poses a challenge due to the absence of labels. To overcome this limitation, we harnessed our developed tool, "Payload-Byte" \cite{farrukh2022payload}, to label the raw packets from the CIC-IDS2017 dataset. Payload-Byte utilizes the five-tuple features, including Source IP, Destination IP, Source Port, Destination Port, and Protocol, to match packets with labeled flow-based data instances.

After labeling the packets, our subsequent task was to extract the payload information while discarding extraneous data. Features like IP addresses in the network layer and port fields in the transport layer were omitted due to potential bias and their dynamic nature, rendering them unsuitable for training machine learning models. Consequently, we focused on the payload contents as our primary feature of interest. Given the variable payload sizes in each packet, we standardized their length to a maximum of 1500 bytes, aligning with the common 1500-byte packet size limit \cite{farrukh2022payload}. This extracted payload was then unified into a single comprehensive feature, subsequently divided into 1500 individual features based on bytes. This transformation involved converting the hexadecimal representation of each byte into an integer within the range of 0 to 255, effectively creating a unique feature for every byte. To maintain consistency, we employed zero padding for packets with payloads containing fewer than 1500 bytes, ensuring uniform feature vector structures.
\begin{figure*}[!htbp]
  \centering
  \includegraphics[width=\linewidth]{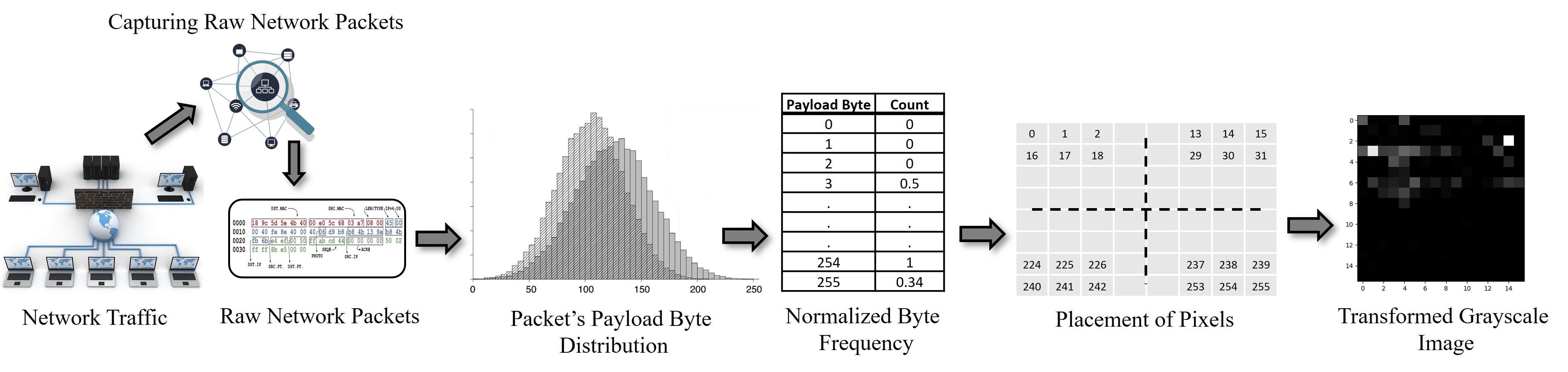}
  \caption{Pictorial representation of the processing pipeline for generating grayscale images from raw network packets. The calculated packet's payload byte distribution is normalized with respect to the highest frequency within each packet.}
  \vspace{-1.4em}
  \label{fig:pipeline}
\end{figure*}

Additionally, we conducted preprocessing steps as part of our data preparation. This encompassed the removal of duplicate instances and the exclusion of instances lacking payload data. To manage dataset size effectively, we implemented under-sampling, which aided in reducing the number of benign instances. For a more comprehensive understanding of these preprocessing procedures and the functionality of Payload-Byte, we refer interested readers to our prior work \cite{farrukh2022payload}.

Subsequently, we transformed our processed data into grayscale images, leveraging payload byte frequency. The data was then split into three subsets: $D_1$, $D_2$ and $D_3$ with the ratio 50:20:30 respectively. The subset $D_1$ is utilized for training base learner models, whereas $D_2$, intended for the meta-learner model. The subset $D_3$ is set aside for evaluating the proposed approach.

\subsection{Payload Transformation into Images}
Our approach to transforming network traffic packet payloads into grayscale images hinges on the byte frequency distribution within each packet's payload data. This transformation addresses a core challenge in dealing with packet-based network traffic—managing extensive data volumes and complex feature spaces. By converting the data into images, we effectively handle high-dimensional data without the need for intricate feature engineering. An overview of the entire process, starting from raw packet data to image generation, is depicted in Fig. \ref{fig:pipeline}.

The transformation process begins with capturing raw packets and extracting their payload data. Following this, we calculate the byte-wise frequency distribution for each packet. This distribution serves as a unique fingerprint, capturing the frequency of each byte within the payload and providing a distinct representation of the packet's content.

To create meaningful grayscale images, we normalize the byte frequency distribution. Instead of using a generic normalization factor, we adopt a packet-specific approach. This means that each packet's byte frequency distribution is normalized based on the highest frequency observed within that specific packet. This tailored normalization strategy ensures that the resulting grayscale images accurately represent the relative importance of byte occurrences within each packet.

\begin{figure}[!htbp]
  \centering
  \includegraphics[scale=0.45]{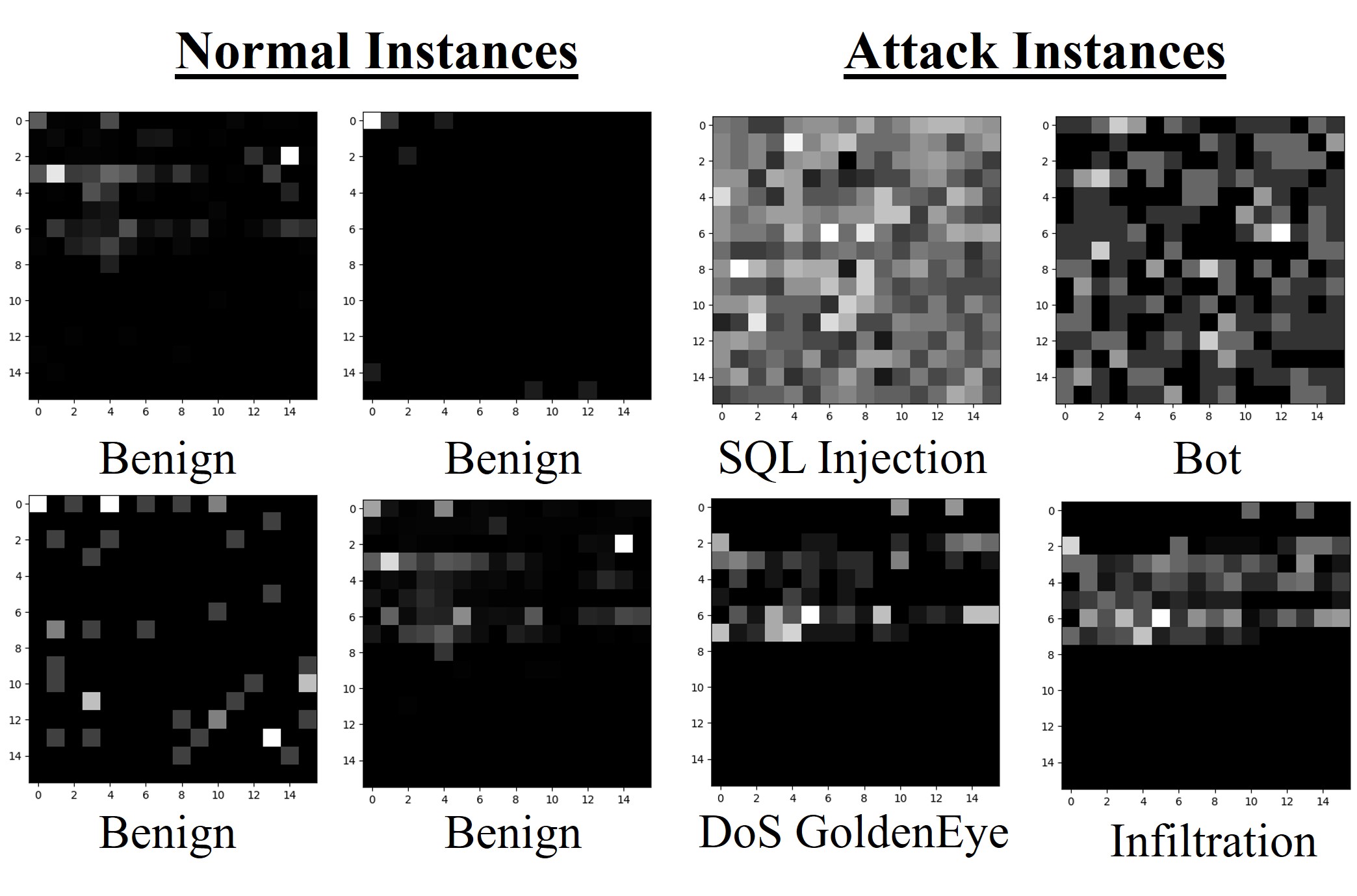}
  \caption{Visualization of grayscale images generated from raw network packets, derived from payload byte frequency distribution. The examples shown are randomly selected for illustration.}
  \vspace{-1.4em}
  \label{fig:example}
\end{figure}

Utilizing the normalized frequency distributions, our next step involves crafting the grayscale images. Each image is designed with dimensions of 16x16 pixels, providing an organized canvas for visualization. Within this 16x16 grid, we linearly arrange the byte frequency values, accommodating all 256 possible byte values. Fig. \ref{fig:example} presents examples of packet-level network traffic after transformation.

\begin{figure*}[!htbp]
  \centering
  \includegraphics[scale=0.118]{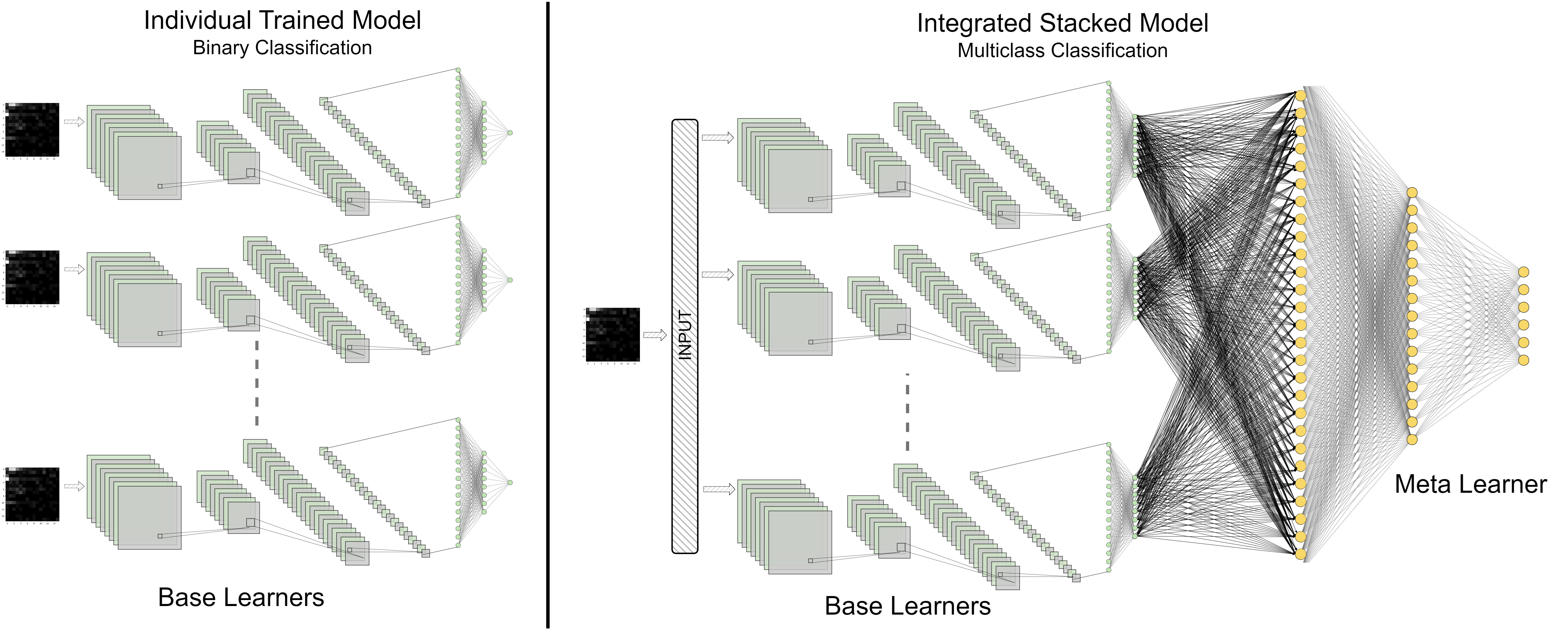}
  \caption{Model architecture of the proposed approach—comprising 15 distinct base learner models, each trained for a specific attack class using a one-vs-all approach. Integrated stacking is achieved through meta learner with additional dense layers integrated into non-trainable concatenated base learners, forming an integrated stacking model.}
  \vspace{-1.4em}
  \label{fig:architecture}
\end{figure*}

\subsection{Integrated Stacked Model}
Our methodology, ByteStack-ID, builds upon the Stacking ensemble technique \cite{stack_2}, which leverages two categories of learners: Base Learners and Meta Learners. Stacking utilizes the classification outputs of these base learners, often referred to as level-0 learners, as inputs for a meta-learner or level-1 learner. This approach excels in handling highly diverse datasets, offering various methods of data classification and typically outperforming standalone classifiers \cite{odegua2019empirical}.

Polikar's foundational work \cite{polikar2006ensemble} outlines four primary strategies for diversifying learning in stacking: using different training data, adjusting training parameters, incorporating diverse features for training, and combining various types of classifiers. In line with this, our ByteStack-ID approach embraces diversity by employing separate training sets for each base learner, treating them as binary classifiers.

Our methodology introduces a unique strategy for the meta-learner. Unlike the conventional approach, where additional models are trained as meta-learners using the outputs of base learners to enhance predictions, we integrate the last layer of each base learner through the meta-learner. This integration creates a larger neural network, treating the stacking ensemble as a unified and cohesive model. Hence, we refer to our approach as an Integrated Stacked Model.

In our approach, each base learner undergoes individual training and is then integrated into additional layers treated as the meta-learner. During the meta-learner's training, the weights for each layer of the base learners are frozen, allowing only the weights of the meta-learner to be learned. This strategic approach significantly enhances the overall performance and adaptability of our intrusion detection model. For a detailed training mechanism, please refer to Algorithm 1.

For a visual representation of ByteStack-ID and its architecture, please see Fig. \ref{fig:architecture}. In this figure, we illustrate 15 distinct base learners, all trained for binary classification, with each base learner addressing a one-vs-all classification problem. After training all the base learners, we integrate them with the meta-learner model layers. Instead of using the last layer of each base learner model, we employ the second last layer for integration with the meta-learner. Subsequently, the integrated model or meta-learner is trained using the $D_2$ sub-dataset.

\begin{figure}[!htbp]
  \centering
  \includegraphics[width=\linewidth]{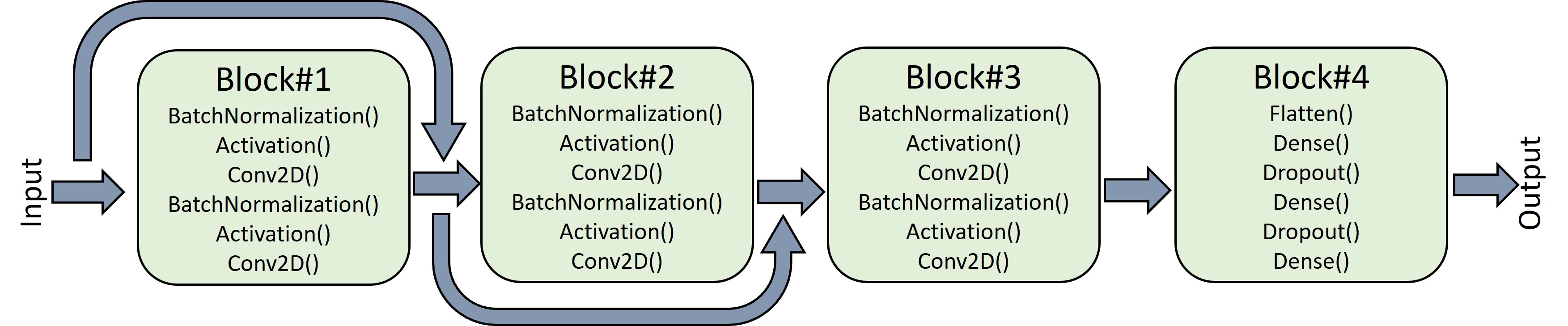}
  \caption{Model architecture of the adopted deep concatenated 2D-CNN for base learners—consisting of four blocks with input and output concatenation for enhanced feature extraction.}
  \label{fig:model}
\end{figure}

\subsubsection{Architecture of Base Learner}

For our base learners, we have built upon the architecture previously introduced in our prior work \cite{farrukh4370422senet}, which leverages the concept of input and output concatenation. The proposed model employs a combination of multiple Convolutional Neural Networks (CNNs) in a cascaded or concatenated manner, enabling the model to acquire insights at multiple levels of abstraction and representations from the input data. Our model architecture aligns with the established 2D-CNN baselines. As per common practice, the number of filters in our proposed model progressively increases across layers, empowering the network to capture increasingly complex and abstract features. You can find a detailed layer-wise progression within the model in Table \ref{tab:architecture}.

\begin{table}[]
\centering
\caption{Detailed Model Architecture Specifications for Base Learners}
\label{tab:architecture}
\resizebox{200pt}{!}{%
\begin{tabular}{|c|c|}
\hline
\multicolumn{1}{|l|}{}             & \textbf{Layer Type}                             \\ \hline
\textbf{Input}                     & InputLayer (16, 16, 1)                          \\ \hline
\multirow{6}{*}{\textbf{Block\#1}} & BatchNormalization(), Activation('Relu')        \\ \cline{2-2} 
                                   & Conv2D(8, (3,3), padding='same')                \\ \cline{2-2} 
                                   & BatchNormalization(), Activation('Relu')        \\ \cline{2-2} 
                                   & Conv2D(8, (3,3), padding='same')                \\ \hline
\multirow{6}{*}{\textbf{Block\#2}} & BatchNormalization(), Activation('Relu')        \\ \cline{2-2} 
                                   & Conv2D(16, (3,3), padding='same')               \\ \cline{2-2} 
                                   & BatchNormalization(),Activation('Relu')         \\ \cline{2-2} 
                                   & Conv2D(16, (3,3), padding='same')               \\ \hline
\multirow{6}{*}{\textbf{Block\#3}} & BatchNormalization(),Activation('Relu')         \\ \cline{2-2} 
                                   & Conv2D(32, (3,3), padding='same')               \\ \cline{2-2} 
                                   & BatchNormalization(), Activation('Relu')        \\ \cline{2-2} 
                                   & Conv2D(32, (3,3), padding='same',strides=(2,2)) \\ \hline
\multirow{6}{*}{\textbf{Block\#4}} & Flatten()                                       \\ \cline{2-2} 
                                   & Dense(64, activation=Activation('Relu'))        \\ \cline{2-2} 
                                   & Dropout(rate=0.2)                               \\ \cline{2-2} 
                                   & Dense(32, activation=Activation('Relu'))        \\ \cline{2-2} 
                                   & Dropout(rate=0.2)                               \\ \cline{2-2} 
                                   & Dense(16, activation=Activation('Relu'))        \\ \hline
\textbf{Output}                    & Dense(1, activation='Sigmoid')                  \\ \hline
\end{tabular}%
}
\vspace{-0.7 em}
\end{table}

Furthermore, our model incorporates a distinctive structural element, wherein the output of block 1 is fused with the input data, effectively forming the input for block 2. This process is iterated for block 3, where the output of block 2 is concatenated with the input of block 2. However, it's worth noting that this same process is not repeated for block 4. This deviation is due to the nature of block 4, which relies on dense layers rather than convolutional layers. Since dense layers operate on flattened feature representations, concatenating the images at this stage would not yield significant benefits in terms of model performance. A pictorial representation of the model is illustrated in Fig. \ref{fig:model}

\begin{algorithm*}
\caption{Training of ByteStack-ID}
\label{algorithm:framework}
\begin{algorithmic}[1]
\Statex \textbf{Input:} Raw Network Traffic Packet
\Statex \textbf{Output:} Multiclass Classification $\rightarrow$ Attack Class
\Statex \textbf{Step 1: Dataset Division}
\State $D_1$: Base Learner Dataset $\leftarrow$ 50\% of the Dataset
\State $D_2$: Meta Learner Dataset $\leftarrow$ 20\% of the Dataset
\State $D_3$: Evaluation Dataset $\leftarrow$ 30\% of the Dataset
\Statex \textbf{Step 2: Packets into Grayscale Images}
\State Extract payload from each packet and calculate byte-wise frequency distribution
\State Normalize frequency distribution using the highest frequency observed within each packet
\State Generate grayscale images with dimensions of 16x16 pixels
\Statex \textbf{Step 3: Base Learner Model Training}
\For{$i = 1$ to $N$} \Comment{$N$ = Number of Attack Classes}
    \State Transform labels of $D_1$ into one-vs-all labels \Comment{1 = Specific Attack Class; 0 = Rest of the classes}
    \State Train the $i_{th}$ base learner model using the transformed $D_1$
    \State Freeze the weights of all layers
\EndFor
\Statex \textbf{Step 4: Meta Learner Model Training}
\State Remove the last layer of each base learner model
\State Integrate the meta learner dense layers with the non-trainable layers of all base learner models
\State Train the integrated model using the $D_2$ dataset
\end{algorithmic}
\end{algorithm*}

\subsubsection{Architecture of Meta Learner}
Our adoption of integrated stacking simplifies the meta-learner's architecture. Instead of introducing an additional model as a meta-learner, which typically trains on the outputs of base learners, our approach directly integrates the dense layers of the meta-learner with the second last layer of each base learner. This choice is accompanied by freezing the layers of the base learner, ensuring that only the weights of the meta-learner layers are trainable. When we apply the $D_2$ sub-dataset to the integrated model, the training process exclusively targets the meta-learner layers.

\begin{figure}[!htbp]
  \centering
  \includegraphics[width=\linewidth]{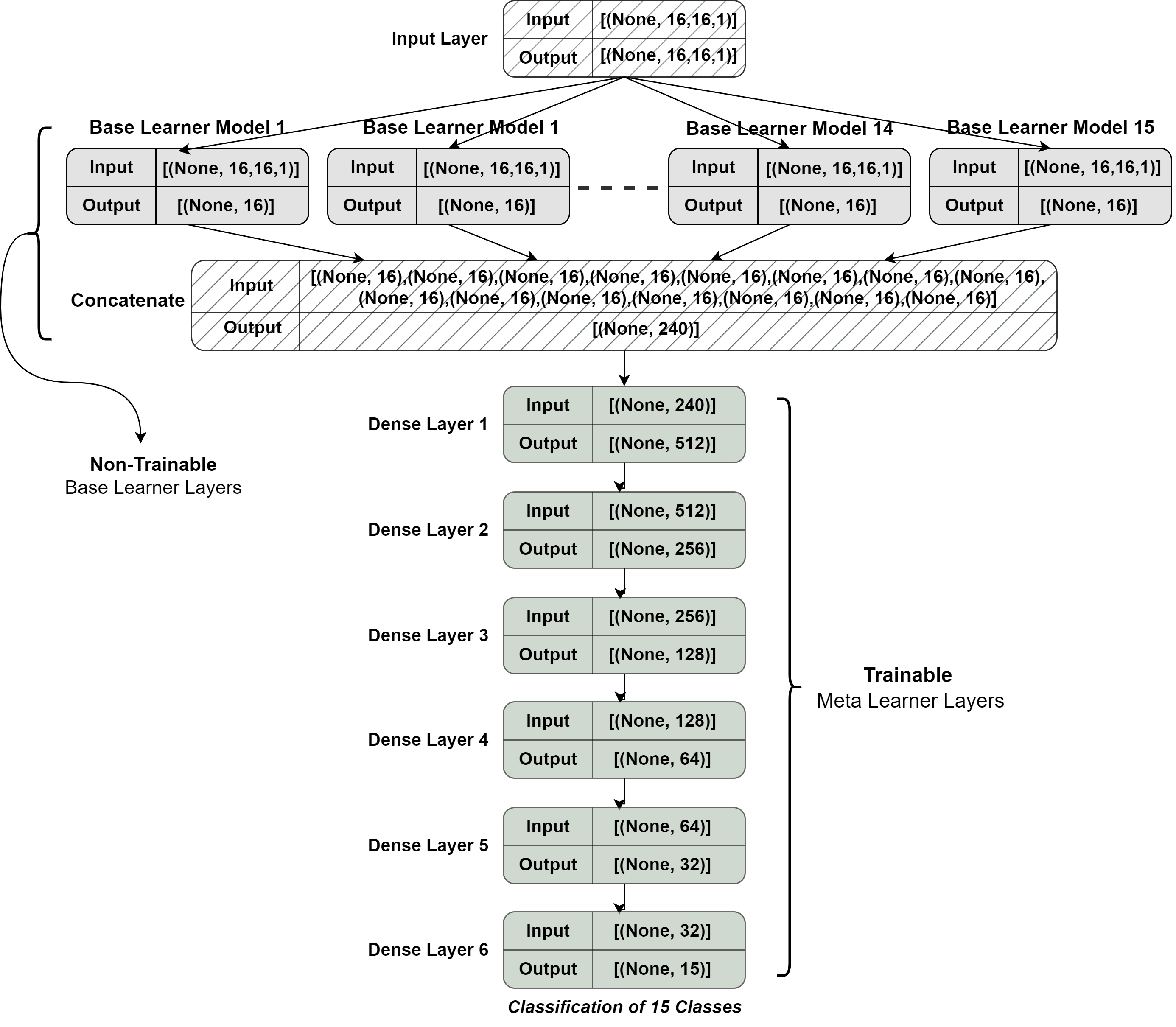}
  \caption{Model architecture of the integrated meta learner and its integration with base learners. Green blocks represent additional layers of the meta learner integrated with all base learners, with base learners' weights frozen as non-trainable.}
  \vspace{-1.4em}
  \label{fig:meta_learner}
\end{figure}

For a visual representation of the Integrated Meta Learner, please refer to Fig. \ref{fig:meta_learner}. This diagram provides a clear illustration of the integration process within our approach. Furthermore, the comprehensive training process of our approach is outlined in Algorithm \ref{algorithm:framework}.

\section{Results and Discussion}

To evaluate the effectiveness of our proposed \textit{ByteStack-ID} approach, we conducted comprehensive experiments comparing its performance with several baseline models using packet-level data. Our model comparison included well-established algorithms such as Random Forest (RF), Logistic Regression, AdaboostClassifier, a Deep Neural Network (DNN) with three layers, and a combination of 1D-CNN and Long Short-Term Memory (LSTM) with five layers. Additionally, we included a base model with a similar architecture to \textit{ByteStack-ID}, but it underwent a single training iteration, unlike our two-step base and meta learner training. This inclusion allowed us to investigate the impact of integrated stacking on model performance.

The outcomes of our experiments are presented in Table \ref{tab:result}, where we evaluated each model's performance based on key metrics, including macro F-1 score, precision, and recall. By analyzing the results, it is apparent that the base model, utilizing transformed grayscale images, outperformed all other baseline models. Nevertheless, our \textit{ByteStack-ID} approach consistently achieved superior results across all evaluated metrics, underscoring its effectiveness. 


\begin{table}[]
\centering
\caption{Performance comparison with baseline models}
\label{tab:result}
\resizebox{200pt}{!}{%
\begin{tabular}{|c|c|c|c|}
\hline
\textbf{Models}       & \textbf{Precision} & \textbf{Recall} & \textbf{F1-Score} \\ \hline
Random Forest         & 62\%               & 56\%            & 58\%              \\ \hline
Logistic   Regression & 48\%               & 42\%            & 42\%              \\ \hline
AdaboostClassifier    & 37\%               & 54\%            & 40\%              \\ \hline
DNN                   & 56\%               & 51\%            & 52\%              \\ \hline
CNN\_LSTM             & 61\%               & 56\%            & 56\%              \\ \hline
Base Model            & 81\%               & 75\%            & 78\%              \\ \hline
ByteStack-ID          & \textbf{84\%}      & \textbf{79\%}   & \textbf{81\%}     \\ \hline
\end{tabular}%
}
\vspace{-1.4em}
\end{table}

To further assess the efficacy of our proposed approach, we conducted a comprehensive comparison with state-of-the-art methods employing packet-level data. Our selection criteria prioritized approaches with experiments on the same dataset, reporting precision, recall, accuracy, and F-1 Score. Despite potential differences in input processing and data extraction, our aim was to provide a holistic evaluation of packet-level data methods.

\begin{figure}[htbp]
  \centering
  \includegraphics[width=\linewidth]{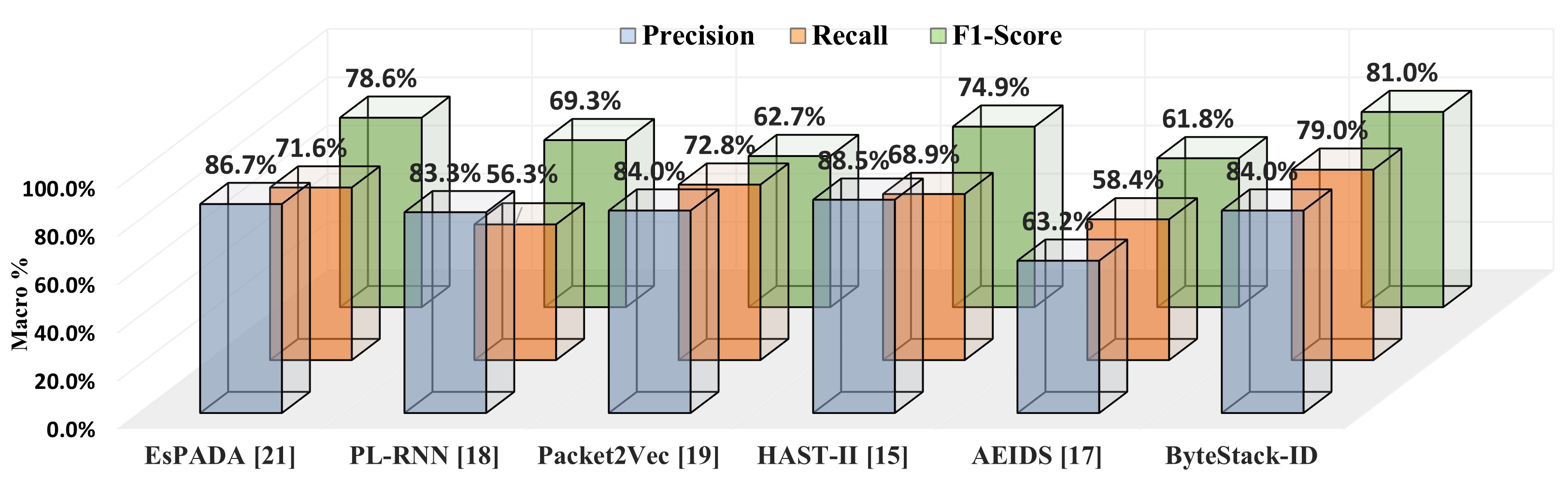}
  \caption{Comparative analysis of the proposed approach against state-of-the-art methods utilizing packet-level data. Performance evaluation utilizes metrics including macro F-1 score, precision, and recall.}
  \label{fig:result}
\end{figure}

To ensure a fair comparison, we identified works under similar experimental conditions.Detailed results in Fig. \ref{fig:result} unequivocally show our approach consistently outperforming other state-of-the-art methods with network traffic packet data. This underscores the effectiveness and superiority of our approach in classifying attack classes using packet-level network traffic data.

\section{Conclusion}

In this paper, we present \textit{ByteStack-ID} a novel method for the NIDS utilizing packet-level network traffic. One of the key innovations of our proposed method lies in the utilization of grayscale images derived from payload data frequency distributions, enhancing the model's ability to detect subtle patterns. ByteStack-ID deviates from conventional stacking methods by integrating additional layers directly into the concatenated base learners, unifying the base learner and meta learner into a single, unified model.

Our empirical findings demonstrate ByteStack-ID's exceptional effectiveness, surpassing both baseline models and state-of-the-art approaches. As network threats evolve, ByteStack-ID proves invaluable in enhancing security measures and proactively mitigating emerging threats within critical network infrastructures. With pioneering techniques and outstanding performance, ByteStack-ID leads intrusion detection research in packet-level network traffic.

\section*{Acknowledgment}


Research reported in this paper was supported by an Early-Career Research Fellowship from the Gulf Research Program of the National Academies of
Sciences, Engineering, and Medicine. The content is solely the responsibility of the authors and does not
necessarily represent the official views of the Gulf Research Program of the National Academies
of Sciences, Engineering, and Medicine.

%



\bibliographystyle{IEEEtran}
\bibliography{references.bib}

\begin{thebibliography}{10}
\providecommand{\url}[1]{#1}
\csname url@samestyle\endcsname
\providecommand{\newblock}{\relax}
\providecommand{\bibinfo}[2]{#2}
\providecommand{\BIBentrySTDinterwordspacing}{\spaceskip=0pt\relax}
\providecommand{\BIBentryALTinterwordstretchfactor}{4}
\providecommand{\BIBentryALTinterwordspacing}{\spaceskip=\fontdimen2\font plus
\BIBentryALTinterwordstretchfactor\fontdimen3\font minus \fontdimen4\font\relax}
\providecommand{\BIBforeignlanguage}[2]{{%
\expandafter\ifx\csname l@#1\endcsname\relax
\typeout{** WARNING: IEEEtran.bst: No hyphenation pattern has been}%
\typeout{** loaded for the language `#1'. Using the pattern for}%
\typeout{** the default language instead.}%
\else
\language=\csname l@#1\endcsname
\fi
#2}}
\providecommand{\BIBdecl}{\relax}
\BIBdecl

\bibitem{10356319}
Y.~A. Farrukh, S.~Wali, I.~Khan, and N.~D. Bastian, ``Detecting unknown attacks in iot environments: An open set classifier for enhanced network intrusion detection,'' in \emph{MILCOM 2023 - 2023 IEEE Military Communications Conference (MILCOM)}, 2023, pp. 121--126.

\bibitem{cite4}
F.~Meneghello, M.~Calore, D.~Zucchetto, M.~Polese, and A.~Zanella, ``Iot: Internet of threats? a survey of practical security vulnerabilities in real iot devices,'' \emph{IEEE Internet of Things Journal}, vol.~6, no.~5, pp. 8182--8201, 2019.

\bibitem{covert}
S.~Wali, Y.~Farrukh, and I.~Khan, ``Covert penetrations: Analyzing and defending scada systems from stealth and hijacking attacks,'' 2024.

\bibitem{hatton2020iot}
M.~Hatton, ``The iot in 2030: Which applications account for the biggest chunk of the 1.5 trillion opportunity?'' 2020.

\bibitem{wali2021explainable}
S.~Wali and I.~Khan, ``Explainable ai and random forest based reliable intrusion detection system,'' 2021.

\bibitem{ghadi2023improved}
Y.~Y. Ghadi, M.~S. Iqbal, M.~Adnan, K.~Amjad, I.~Ahmad, and U.~Farooq, ``An improved artificial neural network-based approach for total harmonic distortion reduction in cascaded h-bridge multilevel inverters,'' \emph{IEEE Access}, vol.~11, pp. 127\,348--127\,363, 2023.

\bibitem{cite5}
\BIBentryALTinterwordspacing
J.~C. Jensen, ``The first workshop on cyber-physical systems education.'' [Online]. Available: \url{https://cps-vo.org/node/7266}
\BIBentrySTDinterwordspacing

\bibitem{farrukh2021sequential}
Y.~A. Farrukh, Z.~Ahmad, I.~Khan, and R.~M. Elavarasan, ``A sequential supervised machine learning approach for cyber attack detection in a smart grid system,'' in \emph{2021 North American Power Symposium (NAPS)}.\hskip 1em plus 0.5em minus 0.4em\relax IEEE, 2021, pp. 1--6.

\bibitem{khan4572172radiant}
I.~Khan, S.~Wali, and Y.~A. Farrukh, ``Radiant: Reactive autoencoder defense for industrial adversarial network threats,'' \emph{Available at SSRN 4572172}.

\bibitem{gupta2020improving}
D.~Gupta and R.~Rani, ``Improving malware detection using big data and ensemble learning,'' \emph{Computers \& Electrical Engineering}, vol.~86, p. 106729, 2020.

\bibitem{mehdi2023machine}
R.~R. Mehdi, M.~Kumar, E.~A. Mendiola, S.~Sadayappan, and R.~Avazmohammadi, ``Machine learning-based classification of cardiac relaxation impairment using sarcomere length and intracellular calcium transients,'' \emph{Computers in Biology and Medicine}, p. 107134, 2023.

\bibitem{syarif2012application}
I.~Syarif, E.~Zaluska, A.~Prugel-Bennett, and G.~Wills, ``Application of bagging, boosting and stacking to intrusion detection,'' in \emph{Machine Learning and Data Mining in Pattern Recognition: 8th International Conference, MLDM 2012, Berlin, Germany, July 13-20, 2012. Proceedings 8}.\hskip 1em plus 0.5em minus 0.4em\relax Springer, 2012, pp. 593--602.

\bibitem{Lee2001RealDetection}
W.~Lee, S.~J. Stolfo, P.~K. Chan, E.~Eskin, W.~Fan, M.~Miller, S.~Hershkop, and J.~Zhang, ``{Real time data mining-based intrusion detection},'' \emph{Proceedings - DARPA Information Survivability Conference and Exposition II, DISCEX 2001}, vol.~1, pp. 89--100, 2001.

\bibitem{farrukh4635437ais}
Y.~A. Farrukh, S.~Wali, I.~Khan, and N.~Bastian, ``Ais-nids: An intelligent and self-sustaining network intrusion detection system,'' \emph{Available at SSRN 4635437}.

\bibitem{wang2017hast}
W.~Wang, Y.~Sheng, J.~Wang, X.~Zeng, X.~Ye, Y.~Huang, and M.~Zhu, ``Hast-ids: Learning hierarchical spatial-temporal features using deep neural networks to improve intrusion detection,'' \emph{IEEE access}, vol.~6, pp. 1792--1806, 2017.

\bibitem{zhang2019multiple}
X.~Zhang, J.~Chen, Y.~Zhou, L.~Han, and J.~Lin, ``A multiple-layer representation learning model for network-based attack detection,'' \emph{IEEE Access}, vol.~7, pp. 91\,992--92\,008, 2019.

\bibitem{AEIDS}
B.~A. Pratomo, P.~Burnap, and G.~Theodorakopoulos, ``Unsupervised approach for detecting low rate attacks on network traffic with autoencoder,'' in \emph{2018 international conference on cyber security and protection of digital services (Cyber Security)}.\hskip 1em plus 0.5em minus 0.4em\relax IEEE, 2018, pp. 1--8.

\bibitem{liu2019cnn}
H.~Liu, B.~Lang, M.~Liu, and H.~Yan, ``Cnn and rnn based payload classification methods for attack detection,'' \emph{Knowledge-Based Systems}, vol. 163, pp. 332--341, 2019.

\bibitem{goodman2020packet2vec}
E.~L. Goodman, C.~Zimmerman, and C.~Hudson, ``Packet2vec: Utilizing word2vec for feature extraction in packet data,'' \emph{arXiv preprint arXiv:2004.14477}, 2020.

\bibitem{Hassan2022IntrusionEmbeddings}
M.~Hassan, M.~E. Haque, M.~E. Tozal, V.~Raghavan, and R.~Agrawal, ``{Intrusion Detection Using Payload Embeddings},'' \emph{IEEE Access}, vol.~10, pp. 4015--4030, 2022.

\bibitem{vidal2020espada}
J.~M. Vidal, M.~A.~S. Monge, and S.~M.~M. Monterrubio, ``Espada: Enhanced payload analyzer for malware detection robust against adversarial threats,'' \emph{Future Generation Computer Systems}, vol. 104, pp. 159--173, 2020.

\bibitem{Sharafaldin2018TowardCharacterization}
I.~Sharafaldin, A.~H. Lashkari, and A.~A. Ghorbani, ``{Toward Generating a New Intrusion Detection Dataset and Intrusion Traffic Characterization},'' 2018.

\bibitem{farrukh2022payload}
Y.~Farrukh, I.~Khan, S.~Wali, D.~A. Bierbrauer, J.~Pavlik, and N.~D. Bastian, ``Payload-byte: A tool for extracting and labeling packet capture files of modern network intrusion detection datasets,'' 2022.

\bibitem{stack_2}
D.~A. Bierbrauer, A.~Chang, W.~Kritzer, and N.~D. Bastian, ``Cybersecurity anomaly detection in adversarial environments,'' \emph{arXiv preprint arXiv:2105.06742}, 2021.

\bibitem{odegua2019empirical}
R.~Odegua, ``An empirical study of ensemble techniques (bagging, boosting and stacking),'' in \emph{Proc. Conf.: Deep Learn. IndabaXAt}, 2019.

\bibitem{polikar2006ensemble}
R.~Polikar, ``Ensemble based systems in decision making,'' \emph{IEEE Circuits and systems magazine}, vol.~6, no.~3, pp. 21--45, 2006.

\bibitem{farrukh4370422senet}
Y.~A. Farrukh, S.~Wali, I.~Khan, and N.~D. Bastian, ``Senet-i: An approach for detecting network intrusions through serialized network traffic images,'' \emph{Available at SSRN 4370422}.

\end{thebibliography}

\end{document}